\documentclass[aps,
		prd,
		reprint,
        superscriptaddress,
        shortbibliography,
		  nofootinbib,
		floatfix,
		notitlepage
		]{revtex4-1}
\pdfoutput=1

\usepackage{tikz-feynman}

\usepackage{amsmath}
\usepackage{amssymb}
\usepackage{slashed}
\usepackage{color}
\usepackage{hyperref}
\hypersetup{colorlinks=true,linkcolor=blue} %%% delete at end

\newcommand{\ud}{{\mathrm{d}}}
\hfuzz=\maxdimen \tolerance=10000
\vfuzz=\maxdimen \tolerance=10000
\newcommand{\LCm}{{\scriptscriptstyle -}}
\newcommand{\LCp}{{\scriptscriptstyle +}}

\newcommand{\LCperp}{{\scriptscriptstyle \perp}}
\newcommand{\be}{\begin{equation}}
\newcommand{\ee}{\end{equation}}
\newcommand{\bi}{\begin{enumerate}}
\newcommand{\ei}{\end{enumerate}}
\newcommand{\sfc}{{\mathsf{c}}}

\newcommand{\sfC}{{\mathsf{C}}}
\newcommand{\sfa}{{\mathsf{a}}}
\newcommand{\sfl}{{\mathsf{l}}}

\begin{document}

\title{Vacuum polarisation effects in impulsive fields}
\author{Anton Ilderton}
\email{anton.ilderton@ed.ac.uk}
\affiliation{Higgs Centre, School of Physics and Astronomy, University of Edinburgh, EH9 3FD, UK}

\author{Hannah Kingham}
\email{h.kingham@sms.ed.ac.uk}
\affiliation{Higgs Centre, School of Physics and Astronomy, University of Edinburgh, EH9 3FD, UK}

\begin{abstract}
Photons impinging on strong electromagnetic fields can change both momentum and helicity state,  due to quantum vacuum polarisation.
We investigate these effects in the collision of photons with impulsive PP-waves, which describe e.g.~the fields of ultra-boosted charge distributions.
We connect our results to vacuum birefringence and quantum reflection in both QED and SUSY QED.
We also compare with helicity flip in plane wave backgrounds, exploring how and when known tree-level relations, relating amplitudes in PP-waves and in plane waves, extend to one-loop corrections.  
\end{abstract}
\maketitle
\onecolumngrid
\section{Introduction}
%%%
The strong-field regime of QED is characterised by the presence of intense electromagnetic fields which lead to non-linear and non-perturbative physical effects~\cite{Fedotov:2022ely}. Recent experiments which collide particles with intense lasers~\cite{Cole:2017zca,Poder:2017dpw,Los:2024ysw}, or aligned crystals~\cite{Wistisen:2017pgr}, have been successful in observing quantum radiation reaction effects in the strong-field regime, while the observation of vacuum polarisation effects are targets of upcoming laser experiments~\cite{King:2015tba,Karbstein:2019oej,Karbstein:2019dxo,Karbstein:2022uwf,Borysov:2022cwc,Macleod:2023asi,Macleod:2024jxl,Ahmadiniaz:2024xob}.
    Strong-field effects can become important in future  colliders~\cite{DelGaudio:2018lfm,Yakimenko:2018kih}, where dense particle bunches generate strong collective Coulomb fields, or may be accessed via novel plasma acceleration schemes~\cite{Jeong:2021isq,Marklund:2022gki}.
    
The most common theoretical approach to `strong-field QED' is the Furry expansion~\cite{Furry:1951zz} in which the coupling of matter to dynamical photons is treated perturbatively, as usual, while the coupling to the strong field, treated as a background, is treated exactly. For this to be practical requires sufficiently symmetric backgrounds~\cite{Heinzl:2017zsr}, or approximations which apply beyond weak coupling~\cite{DiPiazza:2016maj}. New ideas and methods are however still required to go to higher loop and higher multiplicity. This is in order to meet high-precision experimental demands, benchmark simulation schemes, test assumptions, and provide insight into the regime of extremely strong fields where all current methods are conjectured to break down~\cite{RitusRN1,Narozhnyi:1980dc,Fedotov:2016afw,Fedotov:2022ely}.

In this paper we extend recent investigations of QED in `impulsive PP-wave' backgrounds~\cite{Adamo:2021jxz,Copinger:2024twl} to \emph{one-loop} processes and vacuum polarisation effects. We focus throughout on the purely quantum process of scattering, \emph{with} helicity flip, of a photon impinging on the PP-wave. There are several reasons for choosing this particular process and class of fields.
Helicity flip is the microscopic description~\cite{Dinu:2013gaa} of vacuum birefringence~\cite{Toll:1952rq,Heinzl:2006xc}, the result that the vacuum, when exposed to intense electromagnetic fields, behaves like a birefringent medium. Combined with kinematic scattering~\cite{King:2012aw,Sangal:2021qeg} and tailored beams~\cite{Karbstein:2020gzg} in order to optimise the signal-to-noise ratio~\cite{Karbstein:2015xra,Blinne:2018nbd,Gies:2021ymf,Karbstein:2021hwc,Gies:2022pla}, the observation of birefringence has been proposed as a flagship experiement at the European XFEL~\cite{Ahmadiniaz:2024xob}.

Turning to our chosen fields, the class of impulsive PP-waves contains the simplest models of both highly-boosted collective Coulomb fields~\cite{Bonnor:1969rb,Aichelburg:1970dh}, and ultra-short duration vacuum solutions. PP-waves exhibit both structural simplicity (they are supported on a single null hypersurface, characterising their propagation direction) and complexity (they can carry arbitrary spatial geometry, transverse to the propagation direction). This leads to analytically tractable, rich expressions for scattering amplitudes which cleanly expose interesting quantum physics.

PP-waves (in particular shockwaves) have frequently been discussed in connection to eikonal scattering~\cite{Levy:1969cr,Tiktopoulos:1971hi,Eichten:1971kd,tHooft:1987vrq,Dray:1984ha,Jackiw:1991ck,Kabat:1992tb,Adamo:2021rfq}, high-energy QCD~\cite{Balitsky:2001gj,Gelis:2010nm,Caron-Huot:2013fea}, and as probes of causality and transplankian scattering~\cite{Amati:1988tn,Horowitz:1989bv,Verlinde:1991iu,Giddings:2004xy,Lodone:2009qe,Giddings:2010pp,Ciafaloni:2014esa,Camanho:2014apa,Kologlu:2019bco}.  They have been studied rather less frequently in QED than QCD or gravity, see though~\cite{Jackiw:1991ck} for the eikonal and \cite{Tarasov:2019rfp} for deep inelastic scattering. A more systematic investigation of strong-field QED in PP-waves was recently begun, starting with the basic low-multiplicity processes at tree level~\cite{Adamo:2021jxz}, before going to all multiplicity~\cite{Copinger:2024twl}.

Our class of field contains the impulsive limit of a plane wave. Non-impulsive plane waves are frequently employed as models of intense laser pulses~\cite{Gonoskov:2021hwf,Fedotov:2022ely}. Their high degree of symmetry allows for analytic progress in the calculation of observables, but these calculations still become extremely challenging at higher loops and multiplicity. The impulsive limit, on the other hand, offers significant simplifications and a rare potential for obtaining closed form results~\cite{Fedotov:2013uja,Ilderton:2019vot}. (The impulsive limit also lends itself to simulation via quantum computing, see~\cite{Hidalgo:2023wzr}.) Here we will give new closed-form results at one-loop.

This paper is organised as follows. In Sec.~\ref{sec:scattering} we introduce our impulsive PP-wave backgrounds, followed by the dressed spinor wavefunctions and propagator needed for Furry picture calculations. In Sec.~\ref{sec:scattering-calc} we calculate the one-loop amplitude for photon scattering on the PP-wave, with helicity flip. To expose some of the structure in our general result we then turn to examples. In Sec.~\ref{sec:examples-planewave} we consider the plane wave limit, in which  all integrals can be performed to yield closed form expressions.  We consider the infra-red behaviour of our PP-wave amplitude in Sec.~\ref{sec:examples-smallangle}, relating this both to small-angle scattering and the plane wave limit.
        We consider large-angle scattering in Sec.~\ref{sec:examples-largeangle}, and helicity flip in SUSY QED in~\ref{sec:example-SUSY}. We conclude in Sec.~\ref{sec:concs}.

\subsection*{Notation and conventions}
We set $\hbar=c=1$. We work throughout in lightfront coordinates $\ud s^2 = 2\ud x^\LCp \ud x^\LCm - \ud x^\LCperp \ud x^\LCperp$ in which $x^\LCm$ is `lightfront time' while $x^\LCperp=(x^1, x^2)$ are the `transverse' directions. The mass-shell condition is $p^2-m^2 = 2p_\LCp p_\LCm -p_\LCperp p_\LCperp -m^2 = 0$ for mass $m$. We write $n_\mu$ for the the null vector such that $n\cdot x=x^\LCm$. 

\section{Scattering on impulsive PP-waves}\label{sec:scattering}

\subsection{Impulsive pp-waves}
The fields of an impulsive PP-wave
are~\cite{Bonnor:1969rb,Jackiw:1991ck}
\be\label{Fmunu}
    F_{\mu\nu} = \delta(x^\LCm) \big(n_\mu\partial_\nu \Phi(x^\LCperp) - \partial_\mu \Phi(x^\LCperp) n_\nu) \;,
\ee
in which the profile function $\Phi(x^\LCperp)$ is arbitrary. The field is non-zero only on the null hypersurface $x^\LCm=0$. The corresponding current is
\be
    J^\nu(x) = \partial_\mu F^{\mu\nu}(x)
    = n^\nu \delta(x^\LCm) \partial_\LCperp^2\Phi(x^\LCperp) \;.
\ee
For a non-zero current, the PP-wave describes the ultra-boosted (hence pancaked) collective Coulomb fields of arbitrary charge distributions,
encoded in the choice of $\Phi$. For example, writing $r\equiv |x^\LCperp|$, the ultra-boost of a bunch of total charge $Q$ and radius $r_0$ is described by~\cite{Bonnor:1969rb}
\be\label{beam}
    \Phi(x^\LCperp) = \frac{Q}{4\pi}\begin{cases}
        r^2/r_0^2 & r \leq r_0 \\
        1+ \log (r^2/r_0^2)& r  \geq r_0 \;,
    \end{cases}
\ee
see also~\cite{Hollowood:2015elj}.
In the limit $r_0\to 0$ we obtain the shockwave, that is the ultra-boost of the Coulomb field of a single charge.  Another example is 
\be\label{vortex}
    \Phi(x^\LCperp) = x^i H_{ij} x^j \;,
\ee
which corresponds to a homogeneous current if $\text{tr} H\not=0$, while $\text{tr} H=0$ gives a vacuum solution related to electromagnetic vortices~\cite{Bialynicki-Birula:2004bvr} (and gravitational waves~\cite{Zhang:2021lrw}).

We will perform our amplitude calculations for general $\Phi(x^\LCperp)$ -- specific examples will be considered in Sec.~\ref{sec:examples}.  We work throughout with the asymptotically flat potential
\be\label{eq:shock-potential}
    A_\mu(x) = -n_\mu \delta(x^\LCm)\Phi(x^\LCperp) \;.
\ee

\subsection{Scattering with helicity flip}
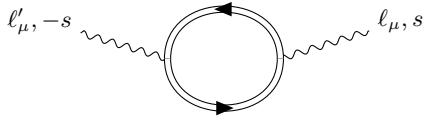
\begin{figure}[t!]
    \centering
\begin{tikzpicture}
\begin{feynman}
    \vertex (a) at (0,0) {$\ell'_\mu,-s$};
    \vertex (b) at (1.7,-0.5);
\vertex[right=of b] (c);
\vertex (d) at (4.8,0.0) {$\ell_\mu,s$};
    \diagram{
    (b) -- [photon] (a);
    (d) -- [photon] (c);
    (c) -- [double,double distance=0.5ex,with arrow=0.5,arrow size=1.5,half right
    ] (b);
    (b) -- [double,double distance=0.5ex,with arrow=0.5,arrow size=1.5,half right] (c);    };
    \end{feynman}
\end{tikzpicture}
\caption{\label{fig:scattering+flip} A photon, momentum $\ell_\mu$ and helicity $s=\pm 1$, collides with a (strong) background field and scatters to momentum~$\ell'_\mu$, and opposite helicity $-s$. The double lines indicate loop particles `dressed', to all orders in perturbation theory, by the background.}
\end{figure}

The process we consider is scattering with helicity flip; a photon, momentum $\ell_\mu$ and helicity $s=\pm 1$, collides with the PP-wave and scatters to momentum $\ell'_\mu$ and helicity $-s$, see Fig.~\ref{fig:scattering+flip}. We will calculate the scattering amplitude in the Furry picture~\cite{Furry:1951zz} (see~\cite{Fedotov:2022ely}for a review), where the coupling of matter to the PP-wave is treated exactly, i.e.~the fermion propagator $S(x,y)$ is a Green function for the Dirac equation in the background (\ref{eq:shock-potential}) -- this is represented by the double line in Fig.~\ref{fig:scattering+flip}. The standard Furry-picture Feynman rules then give the scattering amplitude as
\be\label{raw-amplitude}
    i\mathcal{M} = -(-ie)^2\int \! \ud^4 x\,\ud^4 y\,
    {\bar\varepsilon}^{\,-s}_\mu(\ell')
    e^{i\ell' \cdot x} \;\text{Tr}[\gamma^\mu S(x,y) \gamma^\nu S(y, x)]e^{-i\ell \cdot y} \varepsilon^s_\nu(\ell) \;.
\ee
in which incoming, respectively outgoing, photon wavefunctions which appear in the scattering amplitude are
\be
    e^{-i\ell\cdot y}\varepsilon^s_\nu(\ell) \quad \text{and} \quad  {\bar\varepsilon}^{\,-s}_\mu(\ell')
    e^{i\ell' \cdot x} \;.
\ee
It is convenient to take the photon helicity states $\varepsilon^s_\mu(\ell)$ in the gauge $n\cdot\varepsilon^s=0$. They obey
\be
{\bar \varepsilon}^{\,-s}_\mu(\ell') = \varepsilon^s_\mu(\ell') \;,
\qquad     
\varepsilon^s(\ell)\cdot \varepsilon^s(\ell') = 0 \;,
\qquad
\varepsilon^s(\ell)\cdot \varepsilon^{-s}(\ell') = -1 \;.
\ee
It follows that the null vector $\varepsilon^s_\mu$ actually appears twice in (\ref{raw-amplitude}), greatly reducing the number of possible vector contractions in the amplitude and contributing to the simplicity of later results.

We will construct the propagator $S(x,y)$ from the corresponding scattering wavefunctions in impulsive PP-waves, which are solutions of the Dirac equation in the potential (\ref{eq:shock-potential}). These solutions are obtained simply by patching vacuum solutions across the hypersurface $x^\LCm=0$ where the PP-wave has support, since the potential (\ref{eq:shock-potential}) vanishes everywhere else~\cite{Penrose:1965rx,Klimcik:1988az,Lodone:2009qe}. The (dis)continuity of the solutions is fixed by demanding that the Dirac equation be obeyed \emph{everywhere}, and is expressed through the function
\be\label{W-def}
    W(\sfc) :=\int\!\ud^2 x^\LCperp e^{- i c_\LCperp x^\LCperp + i e \Phi(x^\LCperp)} \;,
\ee
in which we use sans-serif font to refer to purely transverse vectors, so $\sfc=(c_1,c_2)$. In terms of $W(\sfc)$ (the physical interpretation of which we give below) and working in a basis of `lightfront helicity' spin  states~\cite{Chiu:2017ycx}, the wavefunction for an incoming electron, momentum $p$ and helicity label $\sigma=\pm 1$ is~\cite{Adamo:2021jxz}
\be\label{electron-in}
    \psi^\sigma_p(x) =
    \theta(-x^\LCm) e^{-ip\cdot x}u^\sigma_p
    +
    \theta(x^\LCm)\int\!{\hat\ud}^2\sfc\, W(\sfc)\, e^{-i\pi(p,c)\cdot x}u^\sigma_{\pi(p;c)} \;,    
\ee
in which $u_p^\sigma$ are free spinors, while $\pi_\mu(p;c)$ is the on-shell momentum
\be
    \pi_\mu(p;c) = p_\mu - c_\mu + n_\mu\frac{2c\cdot p-c^2}{2n\cdot p} \;,
\ee
with $c_\mu \equiv \delta_\mu^\LCperp c_\LCperp$. What (\ref{electron-in}) says is that as an electron crosses the PP-wave, it is scattered into a superposition of free electron states in which both the electron helicity and the momentum component $n\cdot p = p_\LCp$ are conserved (note $\pi_\LCp = p_\LCp)$, but $p_\LCperp \to \pi_\LCperp = p_\LCperp - c_\LCperp$; the transition amplitude to a particular state is essentially $W(\sfc)$~\cite{Adamo:2021jxz}. The incoming positron wavefunction is, similarly,
\be\label{positron-in}
    \bar\Psi_p(x) = 
     \theta(-x^\LCm) e^{-ip\cdot x}{\bar v}^\sigma_p
    +
    \theta(x^\LCm)\int\!{\hat\ud}^2\sfc\, W^\dagger(\sfc)\, e^{-i\pi(p,-c)\cdot x}{\bar v}^\sigma_{\pi(p;-c)} \;,
\ee
where the change of sign in $\pi$ accounts for the change in sign of the charge.

In terms of the wavefunctions (\ref{electron-in}) and (\ref{positron-in}) the fermion propagator is\footnote{We are working in lightfront coordinates, in which case the propagator really contains an additional `instantaneous' term $\propto \slashed{n}\delta(x^\LCm-y^\LCm)$. This term does not, however, contribute to helicity flip at one loop.} 
\be\label{propagator}
    S(x,y) = \theta(x^\LCm-y^\LCm)\int\!\ud p_{\text{o.s.}} \psi^\sigma_p(x){\bar \psi}^\sigma_p(y)
    -
    \theta(y^\LCm-x^\LCm)
    \int\!\ud p_{\text{o.s.}} \Psi^\sigma_p(x){\bar \Psi}^\sigma_p(y) \;,
\ee
in which repeated spin indices are always summed over, and the on-shell measure in lightfront coordinates is, as usual,
\be
    \int\!\ud p_{\text{o.s.}} = \int\!\frac{\ud^2 p_\LCperp}{(2\pi)^2} \int\limits_0^\infty\!\frac{\ud p_\LCp}{(2\pi)2 p_\LCp} \;.
\ee
The Heaviside theta functions in (\ref{propagator}) imply a natural separation of the amplitude (\ref{raw-amplitude}) into two terms, $\mathcal{M}_>$,  coming from contributions where $x^\LCm > y^\LCm$ and $\mathcal{M}_<$ where $x^\LCm < y^\LCm$.
Using the cyclic properties of the trace it can be shown that these two terms are related by the substitution $\ell \rightarrow -\ell'$ and $\ell' \rightarrow -\ell$, so we proceed by analysing the term 
\be
    i\mathcal{M}_> = -e^2 \int \! \ud^4 x\, \ud^4 y \;e^{i\ell' \cdot x} \theta(x^\LCm-y^\LCm
    ) \text{Tr}[\,
    \slashed{\varepsilon}'
    \!\int\!\ud p_{\text{o.s.}} \psi^\sigma_p(x){\bar \psi}^\sigma_p(y)\,\slashed{\varepsilon}\!
    \int\!\ud q_{\text{o.s.}} \Psi^s_q(x){\bar \Psi}^s_q(y)]e^{-i\ell \cdot y} \;,
\ee
in which, here and below, we have abbreviated $\varepsilon^s(\ell') \to \varepsilon'$, and $\varepsilon^s(\ell)\to \varepsilon$.
Evaluation of this expression benefits from a further split into the regions (i) $0>x^\LCm> y^\LCm$ (ii) $x^\LCm >0>y^\LCm$, and (iii) $x^\LCm>y^\LCm > 0$. In region (i) the field (\ref{Fmunu}) and potential (\ref{eq:shock-potential}) vanish, and $S(x,y)$ reduces to the free propagator -- the calculation of this contribution becomes elementary. In region (iii) the field is also zero, and a change of integration variable from $p\to\pi$ shows that $S(x,y)$ again reduces to the free propagator.
The contribution from region (ii) however is not trivial, as here the fermion propagators straddle the PP-wave. This is 
\be
\begin{aligned}
     i\mathcal{M}_>^{(ii)} = -e^2 \int \! \ud^4x\ud^4y \int \! \ud p_{\text{o.s}}  \ud q_{\text{o.s}}
     \int\!\hat{\ud}^2 \sfc\, \hat{\ud}^2 \sfa \,&
     W(\sfc) W^\dagger(\sfa)
    e^{-i \pi(p, c)\cdot x}e^{ip \cdot y} 
    \;e^{i\ell' \cdot x - i\ell \cdot y}
    \\
    &\times e^{-i \pi(q, -a)\cdot x}e^{iq \cdot y} \text{Tr}[\slashed{\varepsilon}'
    \,u_{\pi(p,c)}^\sigma\bar{u}_p^\sigma\,
    \slashed{\varepsilon}
    \,
    v_q^s\bar{v}_{\pi(q, -a)}^s] \;.
\end{aligned}
\ee
The integrals over $x^\LCp,y^\LCp,x^\LCperp,y^\LCperp$ generate a series of delta functions, which we use to eliminate the $q$ and $a$ integrals. This leaves
\be
\begin{aligned}\label{interacting-incomplete}
   i\mathcal{M}_>^{(ii)} =  -e^2 \int\limits_0^\infty \! \ud x^\LCm &
    \!\int\limits_{-\infty}^0\!\ud y^\LCm
    \!
    \int \! \ud p_{\text{o.s}}  \frac{\theta(\ell_\LCp -p_\LCp)}{2(\ell_\LCp-p_\LCp)} \int\!\hat{\ud}^2 \sfc \; W(\sfc)  \; W^\dagger(\sfc+\sfl' - \sfl)
    \hat\delta(\ell'_\LCp - \ell_\LCp)
    \\ &\times \;
    \exp{\bigg(\frac{-i (\ell' \cdot \pi) x^\LCm + i(\ell \cdot p)y^\LCm}{n \cdot (\ell - p)}\bigg)}
    \text{Tr}\bigg[\,
    \slashed{\varepsilon}'\,
    %(\ell')
    \bigg(1 + \frac{\slashed n \slashed c} {2 n \cdot p} \bigg) (\slashed p + m) \slashed{\varepsilon}
    %(\ell)
    \,(\slashed q -m) \bigg( 1 + \frac{\slashed n \slashed a}{2n \cdot q}\bigg)\bigg] \;,
\end{aligned}
\ee
in which $q$, on-shell, is now given by $q_\mu = \ell_\mu - p_\mu +n_\mu \frac{\ell \cdot p}{n \cdot ( \ell - p)}$.  The trace term is simplified using formulae for e.g.~the trace of six gamma matrices. The lightfront time integrals are performed with appropriate $\pm i \epsilon$ factors to give convergence.
We then find that the full contribution from region (ii) is
\be\label{M-over-2}
   i\mathcal{M}_>^{(ii)} =  -4e^2 \hat\delta(\ell'_\LCp - \ell_\LCp)\int \! \ud p_{\text{o.s}}  \theta(\ell_\LCp -p_\LCp) \int \!\hat{\ud}^2 \sfc\, W(\sfc)  \; W^\dagger(\sfc+\sfl' - \sfl)
    (\ell_\LCp - p_\LCp)
    \frac{(\varepsilon' \cdot \pi)(\varepsilon \cdot p)}{(\ell' \cdot \pi)(\ell \cdot p)} \;.
\ee
The evaluation of the remaining contributions to $\mathcal{M}_>$, from regions (i) and (iii), proceed largely in the same way. The two main differences are in the form of the lightfront time integrands, see immediately below, and in the dependence on $W(\sfc)$. In region (iii) for example, the amplitude contribution cannot depend on the field, but the wavefunctions depend explicitly on $W(\sfc)$. This dependence drops out, as it must, through the identity
\be\label{W-result}
    \int\!\hat{\ud}^2c_\LCperp W(\sfc) W^\dagger(\sfc+\sfa) = \hat{\delta}^2(\sfa) \;,
\ee
which is easily verified using the definition (\ref{W-def}). Adding to (\ref{M-over-2}) the contributions from regions (i) and (iii) we obtain
\be\begin{aligned}\label{integrals-all-regions}
    i\mathcal{M}_> &= 4e^2\; \hat\delta(\ell'_\LCp - \ell_\LCp) \int\!\ud p_{\text{o.s.}} \theta(\ell_\LCp - p_\LCp)(\ell_\LCp - p_\LCp) \\
    &\times\bigg(\int\!
    \hat{\ud}^2 \sfc\, W(\sfc) W^\dagger(\sfc+\sfl' - \sfl)\,
    \frac{\varepsilon \cdot p}{\ell \cdot p}
    \bigg[
     \underset{(\mathrm{iii})}{\underbrace{
         \vphantom{\int\limits_0}\frac{\varepsilon \cdot p}{\ell \cdot p}}}
         -
    \underset{(\mathrm{ii})}
    {\underbrace{
    \vphantom{\int\limits_0}
    \frac{\varepsilon' \cdot \pi}{\ell' \cdot \pi} }}
    \bigg]
    -i\hat{\delta}^2(\sfl'-\sfl)\frac{(\varepsilon \cdot p)^2}{\ell \cdot p}
    \bigg[
    \underset{(\mathrm{iii})}{\underbrace{\int\limits^\infty_0 \ud x^\LCm}}
    + 
    \underset{(\mathrm{i})}{\underbrace{\int\limits_{-\infty}^0\ud x^\LCm }}
   \bigg] \bigg) \;.
\end{aligned}
\ee
There are two points to address. First, note that all terms feature the step function $\theta(\ell_\LCp-p_\LCp)$. This means the second contribution to the amplitude, $\mathcal{M}_<$, will contain the factor $\theta(-\ell_\LCp-p_\LCp)$, but this has no support because $\{\ell_\LCp,p_\LCp\}>0$, so $\mathcal{M}_< = 0$. Hence $\mathcal{M}=\mathcal{M}_>$ and we have found all contributions to the amplitude. Second, consider the final pair of terms in (\ref{integrals-all-regions}). These come from regions (i) and (iii) where the field vanishes, combine to give a divergent volume, and are the terms which (appear to) survive if we turn the PP-wave off. This is exactly the vacuum amplitude -- note that the volume factor is simply the expected fourth momentum-conserving delta function $\hat\delta(\ell'_\LCm - \ell_\LCm)$, which appears as $\hat\delta(0)$ since we already have, in these terms, conservation of three on-shell momenta. We can subtract the vacuum piece at this stage, which imposes the standard on-shell renormalisation condition that the free vacuum polarisation diagram vanishes on-shell. Alternatively, one can impose a regulator to make the $\ud x^\LCm$-integral finite, and proceed to evaluate the loop integrals -- the result vanishes since there is no vector in the integrand which contracts non-trivially with $\varepsilon_\mu$. Either way, we arrive at the following expression for the scattering+flip amplitude; writing $\Delta \ell_\mu \equiv \ell_\mu'-\ell_\mu$ from here on for compactness,
\be\label{shock-amp-all-int-done}
\begin{split}
   i\mathcal{M} &= \hat\delta(\Delta\ell_\LCp)\int\hat{\ud}^2 \sfc\,  W(\sfc) W^\dagger(\sfc+\Delta\sfl)  \, {\widetilde{\mathcal{M}}}(c,\Delta\ell) \;,\\
    \text{where} \quad {\widetilde{\mathcal{M}}}(c,\Delta\ell) & := 4e^2  \int\!\ud p_{\text{o.s.}}
     \theta(\ell_\LCp-p_\LCp) (\ell_\LCp- p_\LCp)
    \bigg[
    \frac{\varepsilon'\cdot p\,\,\varepsilon\cdot p}{\ell'\cdot p \,\, \ell \cdot p}
    -
    \frac{\varepsilon'\cdot \pi\,\,\varepsilon\cdot p}{\ell'\cdot \pi\,\, \ell\cdot p }
    \bigg] \;.
\end{split}
\ee
Written in this form, it is clear that the amplitude vanishes as it should if the background is turned off and we return to vacuum, for then $\pi\to p$. 

%%%%%%%%%%%%%%%%%%%%%%%%%%%%%
\subsection{Evaluation of the integrals} \label{sec:scattering-calc}
%%%%%%%%%%%
%
We proceed to evaluate the loop integrals in (\ref{shock-amp-all-int-done}) as far as possible. For the first term in square brackets of (\ref{shock-amp-all-int-done}) the $c$-integrals can be performed immediately using (\ref{W-result}), which generates a delta function simplifying the remaining dependence on $\ell'$. The loop integrals can then be evaluated exactly and give zero. Turning, then, to the second term in (\ref{shock-amp-all-int-done}), we combine the denominators using the Feynman trick, writing
\be
    \frac{1}{\ell'\cdot\pi \ell\cdot p} = \int\limits_0^1\!\ud x\, \frac{1}{(x \ell'\cdot \pi +(1-x)\ell \cdot p)^2} \;.
\ee
Completing the square in the denominator tells us that the loop integral is simplified by changing integration variables from $(p_\LCperp,p_\LCp)$ to $(P_\LCperp,P_\LCp)$ defined by
\be
\begin{split}
    P_\LCperp &= p_\LCperp - x c_\LCperp - \frac{p_\LCp}{\ell_\LCp} \big( x\, \ell'_\LCperp +(1-x) \ell_\LCperp \big) \;, \\
    P_\LCp &= p_\LCp \;.
\end{split}
\ee
As this corresponds to a \emph{lightlike boost} of the on-shell variable $p_\mu$, the Jacobian for the change of variable is unity. We evaluate the $P_\LCperp$--integral in polar coordinates. The angular integral removes linear factors of $\varepsilon_\LCperp P_\LCperp$ from the numerator of the integrand by symmetry. One then arrives at, writing $P\equiv |P_\LCperp|$,
\be
\begin{aligned}
    \widetilde{\mathcal{M}} = \frac{2e^2 \ell_\LCp}{\pi^2}\int\limits_0^1 \!\ud x\, \int\limits_0^1 \ud u \;u(1- u) \int_0^{\infty} \! \ud P {P}\, \frac{(\varepsilon_\LCperp C_\LCperp)^2 x{(1-x)}}{(P^2 + m^2 + x(1-x) \sfC^2)^2 }
\end{aligned}
\ee
in which this, and subsequent, expressions are simplified by working with the `lightfront momentum fraction' $u:=p_\LCp/\ell_\LCp \in (0,1)$ in the loop, and
\be\label{C-def}
    \sfC\equiv \sfC(u) = \sfc + u \Delta\sfl \;,
\ee
which has the appearance of a total transverse momentum transfer, being the sum of the momentum transfers $\sfl'-\sfl$ to the photon  and $\sfc$ to the massive particles in the loop. The radial integral over $P$ is clearly trivial, and we obtain
\be
\begin{aligned}
     \widetilde{\mathcal{M}} =\frac{e^2 \ell_\LCp}{\pi^2} \int\limits_0^1 \!\ud x\, \int\limits_0^1\! \ud u\, u(1- u)\, \frac{(\varepsilon_\LCperp C_\LCperp)^2 \,x{(1-x)}}{m^2 + x(1-x) \sfC^2} \;.
\end{aligned}
\ee
Some {algebraic manipulation is now required to reduce the  $x$-integral to a recognisable form}. Expanding in partial fractions gives
\be\label{int-tillf}
    \int_0^1\! \ud x \frac{x{(1-x)}}{m^2 + x(1-x)\sfC^2}
    =
    \frac{1}{\sfC^2}\int_0^1\! \ud x \bigg(1 - \frac{m^2}{-\sfC^2(x- \frac{1}{2})^2 + \tfrac{1}{4}\sfC^2+m^2}   \bigg) \;,
\ee
upon which we apply the substitution $z = |\sfC|(x-\frac{1}{2})$. Exploiting the even parity of the integrand to simplify the integration bounds, the integral (\ref{int-tillf}) becomes
\be
    %-
    \frac{2}{|\sfC|^3}\int_{0}^{\frac{|\sfC|}{2}}\! \ud z \bigg(1 - \frac{m^2}{m^2+ \tfrac{1}{4}\sfC^2 - z^2}   \bigg) \;,
\ee
in which the second term is now recognisably the derivative of arctanh.
Evaluating the integral, we finally obtain
\be\label{H-integrals-done}
    {\widetilde{\mathcal{M}}}(c,\Delta\ell) = \frac{e^2\ell_\LCp}{2\pi^2}\int\limits_0^1\!\ud u\, u (1-u) \frac{({C}_1 - i s {C}_2)^2}{\sfC^2}\,
    \Bigg(
    1-
    \frac{4 m^2 \text{arctanh}\bigg[\frac{|\sfC|}{\sqrt{\sfC^2+4 m^2}}\bigg]}{|\sfC|\sqrt{\sfC^2+4 m^2}}
    \Bigg) \;.
\ee
Let us outline some general features of the complete amplitude (\ref{shock-amp-all-int-done}) and (\ref{H-integrals-done}).

First, it can be checked easily that $\widetilde{\mathcal{M}}\to 0$ in the limit that the PP-wave is turned off and we return to vacuum, as it should. (This confirms, as outlined at the start of Sec.~\ref{sec:scattering-calc}, that the first, `free', term in (\ref{shock-amp-all-int-done}) is indeed zero.) Next, only one component of momentum is conserved  so the photon can genuinely scatter, i.e.~change $4$-momentum, as well as flip helicity. This is in contrast to the case of plane waves, discussed in the introduction and below, in which scattering with or without helicity flip is always forward. The function ${\widetilde{\mathcal{M}}}$ is universal, while all the details of the spatial profile $\Phi$ reside in $W(\sfc)$. We will give examples below in which the $u$-integral can be performed analytically. The question of whether or not the $\sfc$-integrals converge depends strongly on the behaviour of $W(\sfc)$, or more precisely on the long-distance behaviour of $\Phi(x^\LCperp)$. Infra-red (IR) divergences should be expected in general, and we will comment more on this below.

We remark that the weak-field limit of our results is obtained by expanding (\ref{W-def}) in powers of $e$. Keeping only terms up to linear order gives
\be\label{W-pert}
    W(\sfc) \to {\hat\delta}^2(\sfc) + ie \widetilde{\Phi}(\sfc) \;,
\ee
in which $\widetilde{\Phi}$ is the Fourier transform of $\Phi$. Inserting this into (\ref{shock-amp-all-int-done}), the $e^0$ term in the product of $W$ factors vanishes, using the result (easily verified) that $\widetilde{\mathcal M}(0,0)=0$, while the linear-in-$e$ terms cancel by symmetry. This leaves
\be\begin{split}
    i\mathcal{M}
    \to
   e^2\hat{\delta}(\Delta\ell_\LCp)
    \int\!\hat{\ud}^2\sfc \,
    {\widetilde \Phi}(\sfc) {\widetilde \Phi}^\dagger(\sfc+\Delta \sfl)\widetilde{\mathcal{M}}(c,\Delta\ell) 
    \;.
\end{split}
\ee
Since $\widetilde{\mathcal M}$ is itself $\mathcal{O}(e^2)$, the whole result is $\mathcal{O}(e^4)$, i.e.~the weak-field limit comes from the four-point one-loop diagram in which two legs are sourced by $\Phi$, as expected in perturbation theory.

%%%%%%%%%%%%%%%%%%%%%%%%%%%%
\section{Examples}\label{sec:examples}
%%%%%%%%%%%%%%%%%%%%%%%%%%%%
%
In this section we investigate the general result above in various limits, and for specific choices of the PP-wave profile~$\Phi$.

%%%%%%%%%%%%%%%%%%%%%%%
\subsection{The plane wave limit}\label{sec:examples-planewave}
We begin with a concrete example in which all integrals can be evaluated explicitly. This will also give a useful comparator problem for subsequent examples, as well as a check on our results so far.

Take the specific PP-wave
\be\label{phi-ipw}
    e\Phi(x^\LCperp) = a_\LCperp x^\LCperp \;,
\ee
in which the factor of $e$ is convention and $a_\LCperp$ is a constant 2-vector. This is the potential for an impulsive plane wave, i.e.~a \emph{vacuum solution} modelling an ultra-short pulse of light. The corresponding $W(\sfc)$ is
\be
    W_{\text{p.w.}}(\sfc) = {\hat \delta}^2(\sfc - \sfa) \;.
\ee
All $\sfc$-integrals in our final results can thus be performed immediately. It is useful to briefly consider the `unevaluated' expression (\ref{shock-amp-all-int-done}); 
the $\sfc$-integrals there generate a transverse delta function setting $\sfl'=\sfl$. Since three components of the on-shell photon momentum are then conserved, we have $\ell'_\mu =\ell_\mu$ exactly. This recovers the result that single photon scattering is always forward in any plane wave background. Expression (\ref{shock-amp-all-int-done}) then becomes 
\be\label{plane-wave-amp}
\begin{split}
   i\mathcal{M}_{\text{p.w.}} = \frac{2e^2  }{\ell\LCp}\hat\delta_{\text{o.s.}}(\ell'-\ell)
   \int\!\ud p_{\text{o.s.}}
     \theta(\ell_\LCp-p_\LCp) (\ell_\LCp- p_\LCp)
    \bigg[
    \frac{(\varepsilon\cdot p)^2}{(\ell\cdot p)^2} -
   \frac{\varepsilon\cdot \pi\,\,\varepsilon\cdot p}{\ell\cdot \pi\,\, \ell\cdot p }
    \bigg] \;,
\end{split}
\ee
where $\pi\equiv\pi(p;a)$ and $\hat{\delta}_\text{o.s.}$ is the on-shell delta function.  We have verified that this expression matches the impulsive limit of the helicity flip  amplitude calculated for \emph{general} plane waves in~\cite{Dinu:2013gaa}: this provides the promised check on our results.  Turning to (\ref{H-integrals-done}), with (\ref{shock-amp-all-int-done}), the final expression for the full helicity-flip amplitude in an impulsive plane wave can be presented in closed form, as the $u$-integral becomes trivial; the result is 
\be\label{T-def}
 i\mathcal{M}_{\text{p.w.}} = \hat\delta_{\text{o.s.}}(\ell'-\ell) \frac{e^2}{24\pi^2} \frac{(a_1 - i s a_2)^2}{\sfa^2} \,
    \Bigg(
    1-
    \frac{4 m^2 \text{arctanh}\Big[\frac{|\sfa|}{\sqrt{\sfa^2+4 m^2}}\Big]}{|\sfa|\sqrt{\sfa^2+4 m^2}}
    \Bigg) \equiv \hat\delta_{\text{o.s.}}(\ell'-\ell) \mathcal{T}(\sfa)\;.
\ee
Factorising out the on-shell delta function is convenient because what remains, $\mathcal{T}(\sfa)$, is simply related to the total flip probability $\mathbb{P}$ by $\mathbb{P} = |\mathcal{T}(\sfa)|^2$ -- this is shown by including a normalised wavepacket, mod-squaring the amplitude, and integrating over final states, see~\cite{Dinu:2013gaa} for details.

\begin{figure}[t!]
\includegraphics[width=0.5\textwidth]{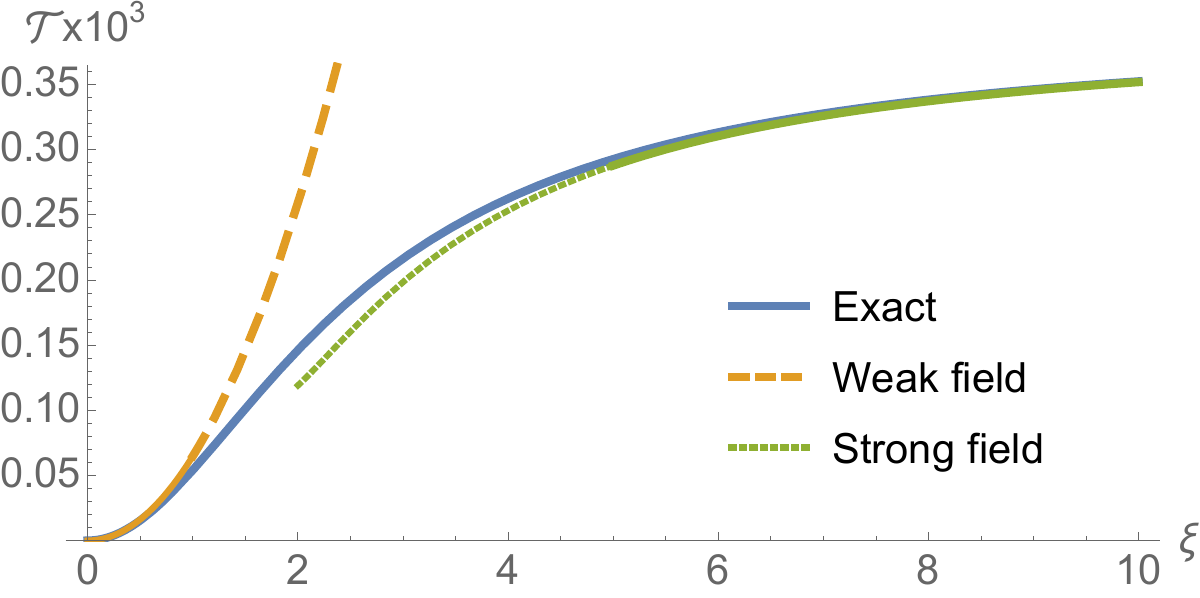}
    \caption{\label{fig:plane-wave} The helicity flip amplitude $\mathcal{T}$ in a plane wave, (\ref{T-planewave})  (solid/blue line) together with its leading order weak (dashed/orange) and strong (dotted/green) field expansions (\ref{T-expand}).}
\end{figure}

Given that all integrals have been evaluated in this plane wave case, it is straightforward to analyse the behaviour of the flip amplitude. To illustrate, take $a_1= m \xi$ and $a_2=0$, corresponding to linear field polarisation, with $\xi$ the `dimensionless intensity parameter'~\cite{Gonoskov:2021hwf,Fedotov:2022ely} characterising the strength of the field. We then have
\be\label{T-planewave}
    \mathcal{T} = 
    \frac{e^2}{24 \pi ^2 \xi }
        \bigg(\xi -\frac{4}{\sqrt{\xi ^2+4}}\, \text{arctanh}\bigg[\frac{\xi }{\sqrt{\xi ^2+4}}\bigg]\bigg) \;.
\ee
We plot this $\mathcal{T}$ in Fig.~\ref{fig:plane-wave}, together with its leading perturabtive ($\xi\ll 1$), and strong field ($\xi\gg 1$) approximations
\be\label{T-expand}
    \mathcal{T}\bigg|_{\xi\ll 1} \simeq \frac{e^2 \xi^2}{144 \pi ^2} \,, \qquad 
     \mathcal{T}\bigg|_{\xi\gg 1} \simeq 
     \frac{e^2}{24 \pi ^2}-\frac{e^2 \log \xi}{6 \pi^2 \xi^2} \;.
\ee
The weak field limit shows the expected behaviour: if we treat the background perturbatively, the first nontrivial contribution to $\mathcal{T}$ comes from 4-photon scattering, in which our incoming and outgoing photons give a factor of $e^2$ while the other two photons are drawn from the background and give a factor $\xi^2$. The strong-field limit $\xi\gg 1$ shows, as well as the  logarithmic dependence on field strength typical of amplitudes in impulsive plane waves~\cite{Ilderton:2019vot}, an explicit upper limit on the flip probability. It would be interesting to see how higher loop corrections change this limit. 

\subsection{Small angle scattering and the infra-red}\label{sec:examples-smallangle}
For our next example we consider the limit of small-angle scattering, such that $ \Delta \ell_\LCperp \simeq 0$; this is equivalent to low momentum transfer and may be thought of as applying a geometric optics approximation to the scattering, in which the incoming photon is undeflected~\cite{Dinu:2014tsa,Hollowood:2015elj}.

One motivation for considering this limit comes from the study of \emph{tree-level} scattering in impulsive PP-waves, where it has been shown that 3-point QED amplitudes can be constructed by taking a weighted average, over all field strengths and polarisations, of the corresponding amplitude in an impulsive \emph{plane} wave~\cite{Adamo:2021jxz}. The weight is $W(\sfc)$. This result extends to \emph{arbitrary} multiplicity photon-emission at tree level (and to arbitrary multiplicity mixed absorption and emission in a certain kinematic regime)~\cite{Copinger:2024twl}. Such results \emph{do not} extend to scattering on gravitational shockwaves and impulsive plane waves, essentially because of the self-interaction of the gravitational field -- this suggests the results may not extend to loop processes in QED, either, where vacuum polarisation effectively introduces photon self-interactions. It is also clear that if the QED results did extend to helicity flip then they could only do so for (near-) forward scattering, since scattering is always forward in the plane wave case, as discussed above. We will address this as part of what follows.

To make the small-angle approximation in the general results (\ref{shock-amp-all-int-done}) and (\ref{H-integrals-done}) requires neglecting the momentum transfer~$\Delta \sfl$ compared to $\sfc$, but whether or not this is allowed depends on how $W(\sfc)$ behaves at small $|\sfc|$: infra-red (IR) problems will typically arise when $\Phi$ does not fall to zero sufficiently quickly at large distance, leading to singularities as the momentum transfer $\sfc\to 0$. For example, in the shockwave proper, $|W(\sfc)|\sim 1/|\sfc|^2$, in which case IR divergences are to be expected.

As an example in which the IR is under control, suppose that $\Phi$ has support in some \emph{finite} area $S$ of the transverse plane, either by construction or by regularisation of an extended profile. It is then possible to isolate the leading IR behaviour of $W(\sfc)$ by adding and subtracting to it the integral of $e^{-ic_\LCperp x^\LCperp}$ over $S$, writing 
\be
\begin{split}
    W(\sfc) &= \int\! \ud^2 x^\LCperp \, e^{-ic_\LCperp x^\LCperp} +
    \int_S\! \ud^2 x^\LCperp \, e^{-ic_\LCperp x^\LCperp}\big(e^{ie\Phi(x^\LCperp)} - 1\big) \\
    &= {\hat \delta}^2(\sfc)  +
    \int_S\! \ud^2 x^\LCperp \, e^{-ic_\LCperp x^\LCperp}\big(e^{ie\Phi(x^\LCperp)} - 1\big) \; \\
    &\equiv  {\hat \delta}^2(\sfc)  + W_{\scriptscriptstyle{\text{IR}}}(\sfc) \;.
\end{split}
\ee
The integral term $W_{\scriptscriptstyle{\text{IR}}}(\sfc)$ is clearly finite as $\sfc\to 0$, hence the leading IR behaviour of $W(\sfc)$ is given by the delta function. The same will hold if $\Phi$ is not compactly supported but falls asymptotically to zero sufficiently quickly. Inserting this into (\ref{H-integrals-done}) and (\ref{shock-amp-all-int-done}) and again using the result that $\widetilde{\mathcal M}(0,0)=0$, we obtain
\be\begin{split}
    i\mathcal{M}
    =
   \hat{\delta}(\Delta\ell_\LCp)
   \bigg(
   W_{\scriptscriptstyle{\text{IR}}}^\dagger(\Delta\sfl)\widetilde{\mathcal{M}}(0,\Delta \ell)
   +
    W_{\scriptscriptstyle{\text{IR}}}(-\Delta\sfl)\widetilde{\mathcal{M}}(-\Delta\ell,\Delta\ell)
  +
    \int\!\hat{\ud}^2\sfc \,
    W_{\scriptscriptstyle{\text{IR}}}(\sfc)W_{\scriptscriptstyle{\text{IR}}}^\dagger(\sfc+\Delta \sfl)\widetilde{\mathcal{M}}(c,\Delta\ell) 
    \bigg) \;.
\end{split}
\ee
Now we can consider the forward scattering limit; setting $\Delta\ell_\perp = 0$ the first two terms vanish, while in the final term the $u$-integral can be performed directly, leaving
\be\begin{split}\label{PP-plane-realation}
    i\mathcal{M}\bigg|_{\ell'_\LCperp=\ell_\LCperp}
= 2\ell_\LCp \delta(\Delta \ell_\LCp) \int\!\hat{\ud}^2\sfc \,
    |W_{\scriptscriptstyle{\text{IR}}}(\sfc)|^2 \, \mathcal{T}(\sfc) \;,
\end{split}
\ee
in which $\mathcal{T}(\sfc)$ is precisely the impulsive plane wave amplitude defined in (\ref{T-def}). Hence the forward scattering + helicity flip amplitude in an impulsive PP wave is given by averaging the helicity flip amplitude in an impulsive plane wave, over all field strengths and polarisations, with weight $|W_{\scriptscriptstyle{\text{IR}}}(\sfc)|^2$.

We have thus found a relation between one-loop amplitudes in PP-waves and plane waves which is not dissimilar to those previously found at tree level. The first difference is that the relation holds only in a restricted kinematic regime. The second difference is the change in weight from linear to quadratic in $W(\sfc)$. We note this is consistent with the worldline calculations of~\cite{Tarasov:2019rfp}. This also seems consistent with the interpretation in~\cite{Adamo:2021jxz} that the PP-wave appears as a stochastic plane wave to the massive particles crossing it; there was a single particle in the tree-level calculations, which required one averaging of $W(\sfc)$, and while there are two particles here, in the loop, momentum conservation means they receive equal and opposite kicks from the PP-wave (represented by $W(\sfc)$ for the electron and  $W^\dagger(\sfc)$ for the positron), hence there is only one average over $|W(\sfc)|^2$.

%%%%%%%%%%%%%%%%%%%%%%%%
\subsection{Wide angle scattering and quantum reflection}\label{sec:examples-largeangle}
%%%%%%%%%%%%%%%%%%%%%%%%
Quantum reflection is the \emph{back}-scattering of photons on spatially inhomogeneous electromagnetic fields~\cite{Gies:2013yxa,Gies:2014wsa}.
In our class of fields, exact back-scattering is typically forbidden, due to the conservation of the null momentum $\ell_\LCp$. To see this consider the simplest setup of a head-on collision between the PP-wave and the photon, which means setting $\ell_\LCperp = \ell_\LCm = 0$. Back-scattering would then require $\ell'_\LCperp=\ell'_\LCp=0$, but this is kinematically forbidden since $\ell'_\LCp=\ell_\LCp\not=0$ is conserved. Away from exact back-scattering there is no kinematic obstruction, although we expect the scattering probability to be suppressed. To investigate this we can conveniently parameterise the emitted photon momentum as
\be
    \ell'_\LCp = \ell_\LCp \;,
    \qquad
    \ell'_\LCm = \frac{\ell_\LCp}{2\eta^2} \;,
    \qquad  
    \ell'_\LCperp = \frac{\ell_\LCp}{\eta}(\cos\phi,\sin\phi) \equiv \frac{\ell_\LCp}{\eta} \mathsf{v}\;,
\ee
in which $\phi$ is the scattering angle in the transverse plane. Here $\eta\in(0,\infty)$ should be thought of as an angular variable; in terms of Cartesian coordinates it is related to the polar scattering angle $\theta$ by
\be
    \eta = \frac{1+\cos\theta}{\sqrt{2}\sin\theta} \;,
\ee
and hence interpolates between forward scattering as $\eta\to\infty$ and back scattering as $\eta\to 0$. We also note the scattered photon energy $\omega$ can be written
\be
 \omega =  \frac{\sqrt{2}\ell_\LCp}{1+\cos\theta} \;,
\ee
and hence wide-angle scattering corresponds to \emph{large} momentum transfer.

Inserting the chosen kinematics into our general results, we can expand (\ref{H-integrals-done}) on the assumption that $W(\sfc)$ falls off for large $|\sfc|$, thus obtaining
\be
    i\mathcal{M}\sim -\frac{e^2}{12\pi^2} e^{-2is\phi}\hat\delta(\ell_\LCp' - \ell_\LCp) \int \!\hat{\ud}^2 \sfc\,
    W(\sfc) W^\dagger(\sfc+\frac{\ell_\LCp}{\eta}\mathsf{v})  \bigg[\ell_\LCp - 6is \eta\,  \epsilon_{ij}{ v}_i c_j  + \mathcal{O}(\eta^2)\bigg] \;,
    \ee
in which $\epsilon_{ij}$ is the alternating tensor and $\epsilon_{12}=1$.

The first term in large brackets is the $\mathcal{O}(\eta^0)$ term of the expansion; there is no $\sfc$-dependence in that factor, which allows us to use (\ref{W-result}) to perform the $\sfc$-integrals, yielding a delta function $\delta^2(\mathsf{v}/\eta)$ which has no support in the considered limit, and so this term vanishes -- this is a good consistency check, as it simply says that for exact back-scattering the amplitude vanishes, as we already knew it must for kinematic reasons.

Turning to the order $\eta$ contribution, consider the integral over $\sfc$;
\be
    {v}_i\epsilon_{ij}\int \!\hat{\ud}^2 \sfc\,  W(\sfc)W^\dagger\Big(\sfc+\frac{\ell_\LCp}{\eta}\mathsf{v}\Big)\, c_j\;.
\ee
Unless $\Phi$, thus $W(\sfc)$, has a strong dependence on particular spatial directions, the vector integral can only generate results proportional to $v_j$, but these will vanish when contracted with the alternating tensor. In this case the amplitude is suppressed to (at least) quadratic order in our expansion; this emphasises the necessity of spatial inhomogeneity for close-to-back scattering. 

\subsection{SUSY QED}\label{sec:example-SUSY}
For our final example, we consider the amplitude for scattering with helicity flip in SUSY QED. It has long been known that in the low energy/weak field Euler-Heisenberg approximation to SUSY QED, the helicity flip amplitude vanishes~\cite{Duff:1979bk,Rebhan:2017zdx}. It was recently shown, though, that the amplitude actually vanishes exactly in arbitrary plane wave backgrounds, for any photon energy, both in SUSY QED and in $\mathcal{N}=1$ SUSY Yang-Mills~\cite{Adamo:2021hno} (see also \cite{Adamo:2019zmk}). This is a slightly surprising result which does not obviously follow from any SUSY ward identity, see~\cite{Adamo:2021hno} for further discussion in relation to helicity conservation in other SUSY processes~\cite{Gounaris:2005ey,Gounaris:2006zm}. It naturally prompts us to ask how the flip amplitude looks, in a PP-wave, in SUSY theories. We will calculate it here for SUSY QED.

To do so we need the contribution to photon helicity flip from a scalar field in the loop, i.e.~the \emph{scalar} QED result. The amplitude is 
\be
% i\mathcal{M}_{\text{scalar}} = e^2\int\!\ud^4x\ud^4y\, \varepsilon^{\prime \mu} e^{i\ell' \cdot x}\overset{\leftrightarrow}{\partial^x_\mu}\, G(x,y)\,    \overset{\leftrightarrow}{\partial^y_\nu}\, G(y,x)e^{-i\ell \cdot y}\varepsilon^\nu
i\mathcal{M}_{\text{scalar}} = e^2\int\!\ud^4x\ud^4y\,  e^{i\ell' \cdot x}\varepsilon^{\prime}\cdot\overset{\leftrightarrow}{\partial_x}\,
G(x,y)\,       \varepsilon\cdot\overset{\leftrightarrow}{\partial_y}\, G(y,x)
    e^{-i\ell \cdot y}
 \;,
\ee
in which the scalar propagator $G(x,y)$ on the PP-wave background is
\be
    G(x,y) =
    \theta(x^\LCm-y^\LCm)\int\!\ud p_{\text{o.s.}} \phi_p(x){\bar \phi}_p(y)
    +
    \theta(y^\LCm-x^\LCm)
    \int\!\ud p_{\text{o.s.}} \Phi_p(x){\bar \Phi}_p(y) \;,
\ee
with $\phi$ and $\bar\Phi$ given by deleting the spinor factors in (\ref{electron-in}) and (\ref{positron-in}) respectively. Note there there is no contribution to the amplitude from the seagull vertex, since $\varepsilon^{s}(\ell')\cdot\varepsilon^s(\ell)=0$, hence we have not written it explicitly.  The calculation of $\mathcal{M}_{\text{scalar}}$ proceeds exactly as in Sec.~\ref{sec:scattering}, so we simply state the final result, which is:
\be
    \mathcal{M}_{\text{scalar}} = -\frac12 \mathcal{M} \;.
\ee
It follows that in SUSY QED, in which two scalars and one spinor run in the loop, the amplitude for scattering with helicity flip vanishes exactly in any impulsive PP-wave background,
\be
    \mathcal{M}_\text{SUSY} = 0\;.
\ee
We thus find that the vanishing of the flip amplitude is not specific to plane waves. It would be interesting to understand more precisely the origin of this result. 

\section{Conclusions}\label{sec:concs}
The collision of particles with impulsive PP-waves presents a largely analytic playground in which to investigate scattering in strong fields. Here we have calculated and studied the one-loop scattering amplitude for photon scattering, with helicity flip, in collision with an impulsive PP-wave. The ultra-short support of the PP-wave lead us to a compact expression for the amplitude, which we investigated in  the limits of low and high momentum transfers.

We also made two connections to helicity flip in plane wave backgrounds. First, the plane wave is a specific example of a PP-wave, and in this case we were able to evaluate all integrals analytically,
giving a closed form expression for the amplitude and the scattering probability. This adds a one-loop result to the small set of closed form scattering amplitudes in strong background fields. Second, we identified an extension, to our loop process, of statements relating tree-level amplitudes in impulsive PP-waves and plane waves: these say that the former is given by a weighted average over the latter, with weight $W$. We saw that a similar structure holds for scattering with helicity flip, except that (i) the weight becomes the IR-safe part of $|W|^2$, and (ii) there is a kinematic restriction, with the relation only holding exactly at forward scattering, due to the special kinematics of $1\to 1$ scattering in a plane wave background. This emphasises the difference between plane waves and PP waves, as the latter allow us to investigate vacuum polarisation effects on both momentum \emph{and} internal degrees of freedom.

Finally, we extended our results to SUSY QED, and found that the helicity flip+scattering amplitude vanishes exactly, for any PP-wave background, and for any photon energy.

PP-waves approximate the fields of ultra-boosted charge distributions, suggesting themselves as natural field models in high-energy accelerator scenarios. Indeed, this is the context in which they are commonly studied, at least in QCD~\cite{Balitsky:2001gj,Gelis:2010nm,Caron-Huot:2013fea}. An interesting question to explore in future work is when this approximation becomes appropriate in the context of future colliders searching for strong field QED effects~\cite{DelGaudio:2018lfm,Yakimenko:2018kih}; in current analyses the dominant theoretical model is based on the `locally constant field approximation'~\cite{Ritus:1985vta} which takes the high field strength -- of e.g.~a dense particle bunch -- to be the dominant scale, rather than the ultra-relativistic velocity of that bunch. Understanding if or when the PP-wave model is more appropriate could potentially simplify predictions and analysis for future QED experiments involving ultra-boosted particles and strong fields. Indeed, understanding this point is one of the longer-term motivations behind~\cite{Adamo:2021hno} and this paper.

It would also be interesting to see if the relation (\ref{PP-plane-realation}) between PP-waves and plane waves has some natural interpretation in terms of eikonal scattering. Further topics for investigation include other one-loop processes, e.g.~electron spin flip, or photon scattering \emph{without} flip, which we did not calculate here, and of course higher-loop processes. It would be intriguing to see how much progress could be made here, in particular in the plane wave case, given current interest in higher loop corrections in strong-field QED~\cite{Fedotov:2016afw,Fedotov:2022ely}, and their contrasting behaviours in different physical regimes~\cite{Podszus:2018hnz,Ilderton:2019kqp,Karbstein:2019wmj,Mironov:2020gbi,Dunne:2021acr,Mironov:2021ohk,Heinzl:2021mji,Torgrimsson:2021wcj,Podszus:2022jia,Torgrimsson:2022ndq}.
\enlargethispage{10pt}
\begin{acknowledgements}
    A.I.~thanks T.~Adamo for useful discussions. The authors are supported by
    the STFC consolidator grant ``Particle Theory at the Higgs Centre,'' ST/X000494/1 (A.I.) and a ``School of Physics and Astronomy Career Development Summer Scholarship'' from the University of Edinburgh (H.K.).
\end{acknowledgements}

\end{document}